\documentclass[showpacs,prd]{revtex4}

\usepackage{epsfig}
\usepackage{graphicx}
\usepackage{amsmath}
\usepackage{latexsym}
\usepackage{color}
\begin{document}

\title{Path integral action and exact renormalization group dualities for quantum systems in noncommutative plane}  
\author{Sunandan Gangopadhyay$^{a,b,c}$\footnote{e-mail: sunandan.gangopadhyay@gmail.com}}
\affiliation{$^a$Department of Physics, West Bengal State University, Barasat, Kolkata 700126, India\\
$^b$Sun Laboratory of Theoretical Physics (SLTP),\\
A7/5, Karunamoyee Housing Estate, Block ED, Sector II, Salt Lake, Kolkata 700091, India\\
$^c$Inter University Centre for Astronomy and Astrophysics (IUCAA), Pune 411007, India\\}









\begin{abstract}
\noindent We employ the path integral approach developed in \cite{sgfgs} to discuss the (generalized)
harmonic oscillator in a noncommutative plane. The action for this system is derived in the coherent state basis 
with additional degrees of freedom. From this the action in the coherent state basis without any additional degrees of freedom is obtained. This gives the ground state spectrum of the system. We then employ the exact renormalization group approach to show that 
{{an equivalence}} can be constructed between this (noncommutative) system and a commutative system. 


\end{abstract}

\maketitle

\noindent The search for a mathematically consistent theory of gravity has made it clear
that such a theory must allow a description in terms of extended objects. This insight first pointed
out in \cite{suss} has led to an intense study of noncommutative quantum mechanics (NCQM) \cite{duval}-\cite{sgthesis}
and noncommutative quantum field theory (NCQFT) \cite{witten}-\cite{carroll} in recent years. A systematic canonical
and path integral formulation of NCQM in a series of papers \cite{fgs}-\cite{sg2} paying careful attention to its interpretational aspects once again revealed the nonlocal character of these theories. These studies further showed the existence of two equivalent pictures
in position representation \cite{rohwer}, namely, a constrained local description in position containing additional degrees of freedom and an unconstrained nonlocal description in position without any other degrees of freedom. The idea of extended objects become transparent from both these descriptions. The connection between the path integral formulations of these two pictures was also demonstrated in
\cite{sg2}.

The idea of noncommutative spacetime has received a new impetus from a very recent study \cite{sgRG} where it has been demonstrated
that dualities can be constructed in a simple quantum mechanical setting using the exact renormalization group (ERG) approach 
\cite{wilson}-\cite{ros} and further such a duality can lead to the emergence of spatial noncommutativity.  
The motivation for this view point originated from the form of the path integral actions derived in \cite{sgfgs,sgfgs1} for a particle in the noncommutative plane without and with a magnetic field. It was noted that these actions differed from the corresponding commutative ones only in a modification of the kinetic energy term, which assumed a simple form when the fields were Fourier transformed.  Hence it was enquired whether these noncommutative actions arose naturally from a generalized ERG approach that only relied on the invariance of the generating functional. The analysis needed the sources to be flowed together with the interacting part of the action. 
This was in contrast to the exact renormalization group equation (ERGE) derived in \cite{pol} which required to preserve the form of the source terms. The entire investigation carried out in \cite{sgRG} involved the quantum mechanical setting of the Landau problem where all computations could be done exactly and analytically.

The aim of this paper is two fold. We consider a (generalized) two dimensional harmonic oscillator \cite{rohwer} in the noncommutative plane. The first part of our investigation is to compute the action for this system using the path integral approach in the coherent state basis $|z; v)$ with additional degrees of freedom. Then using the connection established in \cite{sg2}, the action for this system is obtained in the coherent state basis $|z)$ (which do not have any additional degrees of freedom). It is observed that this action is slightly different from the action for a particle in a magnetic field in the noncommutative plane. The ground state energy is then computed and the results are in agreement with \cite{rohwer} obtained from a canonical analysis. The path integral approach turns out to be {{more elegant and economical than}} the canonical approach. With the action for the system in the noncommutative plane in hand, we then proceed to construct an equivalence between this system and a commutative system employing the generalized ERG approach
developed in \cite{sgRG}. The analysis reveals that the commutative action has to be flowed by modifying the kinetic part of the action
in a way which is different from that adopted in \cite{sgRG}. 

To begin our discussion, we present a brief review of the formalism of 
noncommutative quantum mechanics developed in \cite{fgs}. It was suggested {{there}} that one can give
precise meaning to the concepts of the classical configuration space and the Hilbert space
of a noncommutative quantum system. The first step is to define 
classical configuration space. In two dimensions, 
the coordinates of noncommutative configuration space satisfy the commutation relation 
\begin{equation}
[\hat{x}, \hat{y}] = i\theta
\label{1nc}
\end{equation} 
for a constant $\theta$ that we can take to be positive without any loss of generality. 
The annihilation and creation operators are defined by
$\hat b = \frac{1}{\sqrt{2\theta}} (\hat{x}+i\hat{y})$,
$\hat{b}^\dagger =\frac{1}{\sqrt{2\theta}} (\hat{x}-i\hat{y})$
and satisfy the Fock algebra $[ \hat b , \hat{b}^\dagger ] = 1$. 
The noncommutative configuration space can therefore be viewed as a boson Fock space spanned by the 
eigenstate $|n\rangle$ of the operator $b^{\dagger}b$. This is referred to as the
classical configuration space ($\mathcal{H}_c $)
\begin{equation}
\mathcal{H}_c = \textrm{span}\{ |n\rangle= 
\frac{1}{\sqrt{n!}}(\hat{b}^\dagger)^n |0\rangle\}_{n=0}^{n=\infty}
\label{3}
\end{equation}
where the span is taken over the field of complex numbers.

\noindent The Hilbert space of the noncommutative quantum system is now introduced as
\begin{equation}
\mathcal{H}_q = \left\{ \psi(\hat{x},\hat{y}): 
\psi(\hat{x},\hat{y})\in \mathcal{B}
\left(\mathcal{H}_c\right),\;
{\rm tr_c}(\psi^\dagger(\hat{x},\hat{y})
\psi(\hat{x},\hat{y})) < \infty \right\}.
\label{4hil}
\end{equation}
Here ${\rm tr_c}$ denotes the trace over noncommutative 
configuration space and $\mathcal{B}\left(\mathcal{H}_c\right)$ 
the set of bounded operators on $\mathcal{H}_c$. 
This space has a natural inner product and norm 
\begin{equation}
\left(\phi(\hat{x}, \hat{y}), \psi(\hat{x},\hat{y})\right) = 
{\rm tr_c}(\phi(\hat{x}, \hat{x})^\dagger\psi(\hat{x}, \hat{y}))
\label{inner}
\end{equation}
and forms a Hilbert space \cite{hol}\footnote{States in the noncommutative configuration space are denoted by $|\cdot\rangle$ and 
states in the quantum Hilbert space by $\psi(\hat{x},\hat{y})\equiv |\psi)$ to distinguish between them.}.  
A unitary representation of the noncommutative Heisenberg algebra 
in terms of operators $\hat{X}$, $\hat{Y}$, $\hat{P}_x$ and $\hat{P}_y$ acting on the states of the quantum Hilbert space 
(\ref{4hil}) (assuming commutative momenta) is easily found to be 
\begin{eqnarray}
\hat{X}\psi(\hat{x},\hat{y}) &=& \hat{x}\psi(\hat{x},\hat{y})\quad,\quad
\hat{Y}\psi(\hat{x},\hat{y}) = \hat{y}\psi(\hat{x},\hat{y})\nonumber\\
\hat{P}_x\psi(\hat{x},\hat{y}) &=& \frac{\hbar}{\theta}[\hat{y},\psi(\hat{x},\hat{y})]\quad,\quad
\hat{P}_y\psi(\hat{x},\hat{y}) = -\frac{\hbar}{\theta}[\hat{x},\psi(\hat{x},\hat{y})]~;~\hat{P}=\hat{P}_x +i\hat{P}_y.
\label{action}
\end{eqnarray}
A state (operator) in quantum Hilbert space corresponding to the normalized coherent states (minimal uncertainty states on noncommutative 
configuration space) $|z\rangle$ \cite{klaud} can be constructed as follows
\begin{equation}
|z, \bar{z} )=\frac{1}{\sqrt{\theta}}|z\rangle\langle z|~;~z=\frac{1}{\sqrt{2\theta}}\left(x+iy\right)
\label{csqh}
\end{equation}
and {{satisfies}} 
\begin{eqnarray}
\hat B|z, \bar{z})=z|z, \bar{z})~;~\hat B=\frac{1}{\sqrt{2\theta}}(\hat{X}+i\hat{Y})~,~[\hat B, \hat B^{\ddagger}]=1.
\label{p1}
\end{eqnarray}
{{The notation $\dagger$ is reserved to denote Hermitian conjugation on noncommutative configuration space $\mathcal{H}_c$
and the notation $\ddagger$ for Hermitian conjugation on noncommutative quantum Hilbert space $\mathcal{H}_q$.}}

\noindent The wave-function of a ``free particle" on the noncommutative plane is given by \cite{sgfgs}
\begin{equation}
(z, \bar{z}|p)=\frac{1}{\sqrt{2\pi\hbar^{2}}}
e^{-\frac{\theta}{4\hbar^{2}}\bar{p}p}
e^{i\sqrt{\frac{\theta}{2\hbar^{2}}}(p\bar{z}+\bar{p}z)}
\label{eg3}
\end{equation}
where $|p)$ are the momentum eigenstates given by
\begin{eqnarray}
|p)=\sqrt{\frac{\theta}{2\pi\hbar^{2}}}e^{i\sqrt{\frac{\theta}{2\hbar^2}}
(\bar{p}b+pb^\dagger)}~;~\hat{P}_i |p)=p_i |p)~,~p=p_x +ip_y
\label{eg}
\end{eqnarray}
satisfying the completeness and orthogonality relations
\begin{eqnarray}
\label{eg5a}
\int d^{2}p~|p)(p|&=&1_{q}\\
(p'|p)&=&\delta(p-p')=\delta(p_x-p'_x)\delta(p_y-p'_y).
\label{eg5}
\end{eqnarray}
The completeness relations for the position eigenstates $|z,\bar{z})$ (which is an important
ingredient in the construction of the path integral representation) reads
\begin{eqnarray}
\int \frac{\theta dz d\bar{z}}{2\pi}~|z, \bar{z})\star(z, \bar{z}|=1_{q}
\label{eg6}
\end{eqnarray}
{{where $1_{q}$ is the identity operator on the Hilbert space $\mathcal{H}_q$ of the noncommutative quantum system}} and the star product between two functions $f(z, \bar{z})$ and $g(z, \bar{z})$ is defined as
\begin{eqnarray}
f(z, \bar{z})\star g(z, \bar{z})=f(z, \bar{z})
e^{\stackrel{\leftarrow}{\partial_{\bar{z}}}
\stackrel{\rightarrow}{\partial_z}} g(z, \bar{z})~.
\label{eg7}
\end{eqnarray}
Eq.(\ref{eg6}) can be proved by using eq.(\ref{eg3}) and computing
\begin{eqnarray}
\int \frac{\theta dzd\bar{z}}{2\pi}
(p'|z, \bar{z})\star(z, \bar{z}|p)=
e^{-\frac{\theta}{4\hbar^{2}}(\bar{p}p+\bar{p}'p')}
e^{\frac{\theta}{2\hbar^{2}}\bar{p}p'}\delta(p-p')=(p'|p)~.
\label{eg8n}
\end{eqnarray}
It is possible to decompose the star product $\star=e^{\stackrel{\leftarrow}{\partial_{\bar{z}}}
\stackrel{\rightarrow}{\partial_z}}$ by introducing a further degree of freedom 
\begin{eqnarray}
1_q &=&\int \frac{\theta dz d\bar{z}}{2\pi}~|z, \bar{z})\star(z, \bar{z}|\nonumber\\
&=&\int \frac{\theta dz d\bar{z}}{2\pi}\int dvd\bar{v}~e^{-|v|^2}|z)e^{\bar{v}\stackrel{\leftarrow}{\partial_{\bar{z}}}+
v\stackrel{\rightarrow}{\partial_z}}(z|\nonumber\\
&=&\int \frac{\theta dz d\bar{z}}{2\pi}\int dvd\bar{v}~|z; v)(z; v|
\label{eg8a1}
\end{eqnarray}
where the states $|z; v)$ introduced above are defined by
\begin{eqnarray}
|z; v)&=&e^{-\bar{v}v/2}e^{\bar{v}\partial_{\bar z}}|z, \bar{z})\nonumber\\
&=&e^{\frac{1}{2}(\bar{z}v-\bar{v}z)}|z\rangle \langle z+v|~;~ z, v \in \mathcal{C}
\label{eg8a2}
\end{eqnarray}
and satisfies
\begin{equation}
\hat{B}|z; v)=z|z; v)~;~\forall~ v.
\label{eg8a3}
\end{equation}
For $v=0$, $|z; v)$ reduces to $|z, \bar z)$ given in eq.(\ref{csqh}).

\noindent The overlap of the $|z; v)$ states with the momentum eigenstates $|p)$ reads
\begin{eqnarray}
(z; v|p)=\frac{1}{\sqrt{2\pi\hbar^{2}}}
e^{-\frac{\theta}{4\hbar^{2}}\bar{p}p}
e^{\frac{i}{\hbar}\sqrt{\frac{\theta}{2}}[p\bar{z}+\bar{p}(z+v)]}e^{-\frac{1}{2}\bar{v}v}~.
\label{eg3z}
\end{eqnarray}
We introduce one further notational convention. For any operator $\hat{O}$ acting on the quantum Hilbert space, we may define left and right action (denoted by subscripted $L$ and $R$) as follows:
\begin{eqnarray}
\hat{O}_{L}\psi=\hat{O}\psi~,~\hat{O}_{R}\psi=\psi\hat{O}~;~\forall ~\psi\in\mathcal{H}_{q}.
\label{n1}
\end{eqnarray}
In this language, for instance, the complex momenta $\hat{P}$ may be written as
\begin{eqnarray}
\hat{P}=i\hbar\sqrt{\frac{2}{\theta}}[\hat{B}_{R}-\hat{B}_{L}]~,~\hat{P}^{\ddagger}=i\hbar\sqrt{\frac{2}{\theta}}[\hat{B}^{\ddagger}_{L}-\hat{B}^{\ddagger}_{R}].
\label{n2}
\end{eqnarray}
In this way we have
\begin{eqnarray}
(z; v|\hat{B}^{\ddagger}_{L}|\psi)&=&e^{\frac{1}{2}(\bar{v}z-\bar{z}v)}\langle z|b^{\dagger}\psi|z+v\rangle
=\bar{z}(z; v|\psi),\nonumber\\
(z; v|\hat{B}_{R}|\psi)&=&e^{\frac{1}{2}(\bar{v}z-\bar{z}v)}\langle z|\psi b|z+v\rangle
=(z+v)(z; v|\psi),\nonumber\\
(z; v|\hat{B}_{L}|\psi)&=&e^{\frac{1}{2}(\bar{v}z-\bar{z}v)}\langle z|b\psi|z+v\rangle,\nonumber\\
(z; v|\hat{B}^{\ddagger}_{R}|\psi)&=&e^{\frac{1}{2}(\bar{v}z-\bar{z}v)}\langle z|\psi b^{\dagger}|z+v\rangle.
\label{n3}
\end{eqnarray}
{{The commutation relations of $B_{L}$, $B^{\ddagger}_{L}$, $B_{R}$ and $B^{\ddagger}_{R}$ read $[B_{L}, B^{\ddagger}_{L}]=1=[B^{\ddagger}_{R}, B_{R}]$ and $[B_{L}, B_{R}]=0=[B_{L}, B^{\ddagger}_{R}]$ \cite{rohwer}}}.

\noindent A path integral representation of the propagator in the $|z, v)$-basis
can now be derived in the usual way \cite{klaud, rohwer, sg2} and reads
\begin{eqnarray}
\label{pintegral3}
(z_f; v_f, t_f|z_0;v_0, t_0)&=&N\int \mathcal{D}\bar{\mu}\mathcal{D}\mu~
\exp\left(\frac{i}{\hbar}S\right)\\
\mathcal{D}\bar{\mu}\mathcal{D}\mu&=&\lim_{n\rightarrow\infty}\prod^{n}_{k=1}d\bar{\mu}_{k}d\mu_{k}~;
~d\bar{\mu}_{k}d\mu_{k}=d\bar{z}_{k}dz_{k}d\bar{v}_{k}dv_{k}\nonumber
\end{eqnarray} 
where the action $S$ is given by
\begin{equation}
S=\int_{t_{0}}^{t_{f}}dt~(z; v|i\hbar\partial_{t}-\hat{H}|z; v)~.
\label{action_ncqm}
\end{equation} 
We shall now consider a harmonic oscillator Hamiltonian (augmented by a right action term) and analyze it in the
path integral framework. The Hamiltonian of this reads {{(upto a constant term)}} \cite{rohwer}
\begin{eqnarray}
\hat{H}&=&\frac{1}{2m}\hat{P}\hat{P}^{\ddag}+m\theta\omega_{L}^{2}\hat{B}_{L}^{\ddag}\hat{B}_{L}
+m\theta\omega_{R}^{2}\hat{B}_{R}\hat{B}_{R}^{\ddag}\nonumber\\
&=&\frac{\hbar^2}{m\theta}\left(1+\frac{m^2 \theta^2\omega_{L}^2}{\hbar^2}\right)\hat{B}_{L}^{\ddag}\hat{B}_{L}+\frac{\hbar^2}{m\theta}\left(1+\frac{m^2 \theta^2\omega_{R}^2}{\hbar^2}\right)\hat{B}_{R}^{\ddag}\hat{B}_{R}
-\frac{\hbar^2}{m\theta}(\hat{B}_{L}^{\ddag}\hat{B}_{R}+\hat{B}_{R}^{\ddag}\hat{B}_{L}).
\label{hosc}
\end{eqnarray}  
{{The relevance of this model comes from the fact that this Hamiltonian can also be viewed as a harmonic oscillator with
an added magnetic field. The commutative limit of this model is subtle and shall be discussed subsequently.}}  

\noindent Substituting this form of the Hamiltonian in eq.(\ref{action_ncqm}), we get
\begin{equation}
S=\int dt~\left[i\hbar(\dot{\bar{z}}v-\bar{v}\dot{z}-\bar{v}\dot{v})-m\theta\omega_{R}^{2}(\bar{v}z+\bar{z}v)
-m\theta(\omega_{L}^{2}+\omega_{R}^{2})\bar{z}z-\left(\frac{\hbar^2}{m\theta}+m\theta\omega_{R}^{2}\right)\bar{v}v\right].
\label{a1}
\end{equation} 
Note that this system is a constrained system with the same set of second class constraints \cite{dirac} as in \cite{sgfgs} and 
therefore yields the noncommutative Heisenberg algebra as before \cite{sgfgs} after quantization.

\noindent Setting $\omega_{R}=0$ in the above equation, we obtain the action for the usual harmonic oscillator in the $|z; v)$-basis  
which reads  
\begin{equation}
S=\int dt~\left[i\hbar(\dot{\bar{z}}v-\bar{v}\dot{z}-\bar{v}\dot{v})
-m\theta\omega_{L}^{2}\bar{z}z-\frac{\hbar^2}{m\theta}\bar{v}v\right].
\label{a11}
\end{equation} 
We now proceed to obtain the action in the $|z, \bar{z})$-basis. To compute this, 
we note that the propagator in the $|z, \bar{z})$-basis is related
to the propagator in the $|z; v)$-basis as \cite{sg2}
\begin{eqnarray}
(z_f, \bar{z}_{f}, t_f|z_0, \bar{z}_{0}, t_0)=N e^{-\vec{\partial}_{z_f}\vec{\partial}_{\bar{z}_0}}
\int d\bar{v}dv~(z_f; v, t_f|z_0; v, t_0).
\label{pr1}
\end{eqnarray} 
Setting $v_f =v_0 =v$ in eq.(\ref{pintegral3}), using eq.(\ref{a1}) and integrating over $v$, $\bar v$ leads to
\begin{eqnarray}
\int d\bar{v}dv~(z_f; v, t_f|z_0; v, t_0)=N\int \mathcal{D}\bar{z}\mathcal{D}z~\exp\left(\frac{i}{\hbar}I\right)
\label{pr2}
\end{eqnarray} 
where the action $I$ in the $|z, \bar{z})$-basis is given by
\begin{eqnarray}
I=\int dt~m\theta\left[\frac{1}{\left(1+\frac{m^2 \theta^2\omega_{R}^2}{\hbar^2}\right)}
\left(\dot{\bar{z}}+\frac{im\theta \omega_{R}^2}{\hbar}\bar{z}\right)
\left(1+\frac{im\theta}{\hbar\left(1+\frac{m^2 \theta^2\omega_{R}^2}{\hbar^2}\right)}\partial_t\right)^{-1}\left(\dot{z} -\frac{im\theta \omega_{R}^2}{\hbar}z\right)
-(\omega_{L}^2 +\omega_{R}^2)\bar{z}z \right].
\label{zbas}
\end{eqnarray} 
Interestingly, this action looks similar to the action for a particle in a magnetic field in the presence of a harmonic oscillator
potential moving in the noncommutative plane \cite{sgfgs1} but is not identical to it because of the coefficient $\left(1+\frac{m^2 \theta^2\omega_{R}^2}{\hbar^2}\right)$ 
appearing in the kinetic part of the action. {{The commutative limit of this action can be taken by setting $\theta=0$ in the expression within the parenthesis in the second line of the above equation and absorbing the overall factor $\theta m$ in the measure $dt$ of the integral. It can be seen easily that the action reduces to that of a harmonic oscillator in the presence of a magnetic field in the $\theta=0$ limit
\begin{eqnarray}
I=\int dt~\theta\left[m\dot{\bar{z}}(t)\dot{z}(t)
-m(\omega_{L}^2 +\omega_{R}^2)\bar{z}(t)z(t) \right].
\label{zbasc}
\end{eqnarray} }}
As a consistency check, we once again set $\omega_{R}=0$ in the above equation to obtain the action 
for the usual harmonic oscillator in the $|z, \bar{z})$-basis  
\begin{eqnarray}
I=\int dt~\theta m\left[\dot{\bar{z}}\left(1+\frac{im\theta}{\hbar}\partial_t\right)^{-1}\dot{z}-\omega_{L}^2 \bar{z}z \right]
\label{zbas1}
\end{eqnarray}
which agrees with \cite{sgfgs}. The equation of motion which follows from the action (\ref{zbas}) reads
\begin{eqnarray}
\ddot{z}(t)+\frac{im\theta}{\hbar}(\omega_{L}^2 - \omega_{R}^2)\dot{z}(t)
+\left(\omega_{L}^2 +\omega_{R}^2 +\frac{m^2 \theta^2}{\hbar^2}\omega_{L}^2 \omega_{R}^2\right)z(t)=0.
\label{zeom}
\end{eqnarray} 
{{Note that the non-local action in time (\ref{zbas}) (which owes its origin to the noncommutative parameter $\theta$) gives a local equation
of motion. The effect of the non-locality in the action appears through the presence of terms involving $\theta$ in the equation of motion.}}
Making the ansatz $z_{c}(t)=e^{i\gamma t}$, yield the frequencies
\begin{eqnarray}
\gamma=\frac{1}{2\hbar}\left[-m\theta(\omega_{L}^2 -\omega_{R}^2)\pm\sqrt{(\omega_{L}^2 +\omega_{R}^2)[4\hbar^2 +m^2 \theta^2 
(\omega_{L}^2 +\omega_{R}^2)]}\right].
\label{energy}
\end{eqnarray}
{{We observe that taking $\omega_{+}=\gamma_{+}$ and $\omega_{-}=-\gamma_{-}$ (where $\gamma_{+}$
corresponds to taking the positive sign before the square root in eq.(\ref{energy}) and $\gamma_{-}$
corresponds to taking the negative sign before the square root in eq.(\ref{energy})) agrees with the ground state energy of the system obtained from the canonical approach \cite{rohwer} and also agrees with \cite{sgfgs} in the $\omega_{R}=0$ limit.}} 
The analysis demonstrates the elegance of the path integral approach. This completes our discussion on the first part of the paper, namely, the path integral approach to analyse the (generalized) harmonic oscillator.


We now demonstrate the construction of dualities using the ERG approach \cite{wilson}-\cite{bag}. The analysis shows that one can obtain
{{an action similar to}} the action for the (generalized) harmonic oscillator in the noncommutative plane by flowing the commutative action for a particle in a magnetic field in a harmonic oscillator potential. However, the difference here from the analysis in \cite{sgRG} lies in the choice
of the term modifying the kinetic part of the action. The basic idea of the ERG approach involves the introduction of an ultra-violet cutoff (real) function $K^{-1}(p^2/\ell^2)$ in the theory which has the property that it vanishes when $|p|>\ell$. The equation, known as the Polchinski equation \cite{pol}, is then obtained by requiring that the process of reducing the number of degrees of freedom (commonly known as the coarse graining step) leaves the normalized generating functional invariant. An important ingredient in this derivation, required to preserve the form of the source terms, is the imposition of the condition that the sources vanish above the momentum cutoff and that the cutoff function is independent of the cutoff at small momenta, i.e, $J(p)=0$ for $|p|>\ell$ and $\partial_{\ell}K^{-1}(p^2/\ell^2)=0$ for small $p$ \cite{pol},\cite{banks}. The implication of this is that the effective theory can only yield information on correlation functions of the original theory in as far as they are computed below the momentum cutoff. This approach was extended in \cite{sgRG} by relaxing the conditions imposed on the sources and the cutoff function by requiring only the invariance of the normalized generating functional under the renormalization group flow.  This in turn allowed the computation of all the correlation functions of the original theory in terms of the correlation functions of the effective theory, thereby establishing a complete duality between them.  However,
this requires the flow of the sources together with the interacting part of the action, a view point that has been considered seriously in \cite{ros} to extract correlation functions using the ERGE. 

\noindent We start by summarising the essentials of the ERGE. In our discussion, we consider a complex scalar field theory in 0+1-dimensions, 
the action of which reads
\begin{eqnarray}
\label{action}
S[\phi, \phi^{*}]=\int d\omega~\phi^{*}(\omega)K(\omega, \ell)\phi(\omega)+S_{I}[\phi, \phi^{*}]+J_{\ell}[\phi, \phi^{*}]
\label{1}
\end{eqnarray}
where $K(\omega, \ell)$ takes the standard form $\omega^2$ in the $\ell\rightarrow0$ limit, $S_{I}[\phi, \phi^{*}]$ is the interacting part of the action and $J_{\ell}[\phi, \phi^{*}]$ is a generalized source term which has to be determined by the requirement of invariance of the generating functional.   For our present purposes, involving actions quadratic in the fields, it is simple to see that it is sufficient to limit the form of this functional to be linear, i.e., we take
\begin{eqnarray}
J_{\ell}[\phi, \phi^{*}]=\int d\omega~[J_{0}(\ell) +J^{*}_{0}(\ell) +J_{1}(\ell)\phi^{*}(\omega)+J^{*}_{1}(\ell)\phi(\omega)].
\label{2}
\end{eqnarray}
The initial conditions imposed on $J_{\ell}$ are :
\begin{eqnarray}
\label{2x}
J_{0}(\ell)|_{\ell=0}&=&J^{*}_{0}(\ell)|_{\ell=0}=0 
\label{2y}
\end{eqnarray}
and $J_{1}(0)$ is an arbitrary function of $\omega$ that acts as a source in the bare ($\ell=0$) action. In the subsequent discussion, we denote the first term in eq.(\ref{action}) by $S_{0}[\phi, \phi^{*}]$. 

\noindent The normalized generating functional is given by
\begin{eqnarray}
Z[J_{\ell}]=\frac{\int [d\phi~d\phi^{*}]~e^{-(S_{0}[\phi, \phi^{*}] +S_{I}[\phi, \phi^{*}]+J_{\ell}[\phi, \phi^{*}])}}{\int [d\phi~d\phi^{*}]~e^{-(S_{0}[\phi, \phi^{*}] +S_{I}[\phi, \phi^{*}])}}~.
\label{3}
\end{eqnarray}
We now apply the logic of ERGE as set out in \cite{pol} and \cite{banks}, by requiring this to be invariant under the flow, i.e., it must be independent of $\ell$.  However, in \cite{sgRG}, the conditions imposed in these derivations on $K(\omega, \ell)$ and the sources have been relaxed which in turn requires the flow of the source terms. This leads to the following equations for the interacting part and source terms \cite{pol, sgRG}:
\begin{eqnarray}
\label{9}
\partial_{\ell}S_{I}=\int d\omega~\partial_{\ell}K^{-1}(\omega, \ell)
\left\{\frac{\delta S_{I}}{\delta \phi^{*}(\omega)}\frac{\delta S_{I}}{\delta \phi(\omega)}
-\frac{\delta^{2} S_{I}}{\delta \phi^{*}(\omega)\delta \phi(\omega)}\right\}
\end{eqnarray}
\begin{eqnarray}
\partial_{\ell}J_{\ell}&=&\int d\omega~\partial_{\ell}K^{-1}(\omega, \ell)
\left\{\frac{\delta S_{I}}{\delta \phi(\omega)}\frac{\delta J_{\ell}}{\delta \phi^{*}(\omega)}
+\frac{\delta S_{I}}{\delta \phi^{*}(\omega)}\frac{\delta J_{\ell}}{\delta \phi(\omega)}
+\frac{\delta J_{\ell}}{\delta \phi^{*}(\omega)}\frac{\delta J_{\ell}}{\delta \phi(\omega)}
-\frac{\delta^{2} J_{\ell}}{\delta \phi^{*}(\omega)\delta \phi(\omega)}\right\}.
\label{10}
\end{eqnarray}
Considering the interaction term to be quadratic in the fields, i.e., 
\begin{eqnarray}
S_{I}[\phi, \phi^{*}]=\int d\omega~g(\omega, \ell) \phi^{*}(\omega) \phi(\omega)
\label{11}
\end{eqnarray}
with $g(\omega, \ell)$ real, and using eqs.(\ref{2}) and (\ref{10}) yield the following equations for the source terms
\begin{eqnarray}
\label{13a}
\partial_{\ell}J_{1}(\ell)&=&\partial_{\ell}K^{-1}(\omega, \ell)g(\omega, \ell)  J_{1}(\ell)\\
\partial_{\ell}[J_{0}(\ell)+J^{*}_{0}(\ell)]&=&\partial_{\ell}K^{-1}(\omega, \ell)|J_{1}(\ell)|^{2}.
\label{13c}
\end{eqnarray}
These equations can be integrated (using the initial conditions (\ref{2x}) on the sources) to get
\begin{eqnarray}
\label{14a}
J_{1}(\ell)&=&J_{1}(0)\exp\left(\int_{0}^{\ell} d\ell'~g(\omega, \ell')\partial_{\ell'}K^{-1}(\omega, \ell')\right)\\
J_{0}(\ell)+J^{*}_{0}(\ell)&=&|J_{1}(0)|^{2}\int_{0}^{\ell} d\ell'~\partial_{\ell'}K^{-1}(\omega, \ell')
\exp\left(2\int_{0}^{\ell'} d\ell''~g(\omega, \ell'')\partial_{\ell''}K^{-1}(\omega, \ell'')\right).
\label{14c}
\end{eqnarray}
We now apply these results to the problem of a particle in a magnetic field in a harmonic oscillator potential, the action of which reads
\begin{eqnarray}
S=\int d\omega~[m\bar{z}(\omega)\omega^2 z(\omega) -(eB\omega+k)\bar{z}(\omega)z(\omega) 
+J(\omega)\bar{z}(\omega)+J^{*}(\omega)z(\omega)]
\label{15ha}
\end{eqnarray}
where $k$ is the strength of the harmonic oscillator interaction.  
We now {{let this action flow}} with respect to a parameter $\ell$ as in eq.(\ref{1}), 
using eq.(\ref{11}) for the interaction and taking
\begin{eqnarray}
K(\omega, \ell)=\frac{m\omega^2}{(1-\frac{m\omega\ell}{\hbar}+\frac{eB\ell}{2\hbar})}
\label{17}
\end{eqnarray}
with the initial condition
\begin{eqnarray}
g(\omega, \ell)|_{\ell=0}=-(eB\omega+k).
\label{18}
\end{eqnarray}
The choice of $K(\omega, \ell)$ is motivated by the non-local form of the kinetic energy term appearing in the action for the particle moving in a noncommutative plane in the presence of a (generalized) harmonic oscillator potential (\ref{zbas}). 
{{This can be seen by writing down the action (\ref{zbas}) in the Fourier space :
\begin{eqnarray}
S=\int d\omega~\theta\left[\bar{z}(\omega)P(\omega, \theta)z(\omega)+\frac{m}{(1-\frac{m\omega\theta}{\hbar}+\frac{m^2 \theta^2 \omega^{2}_{R}}{\hbar^2})}\left((\omega^{2}_{L}-\omega^{2}_{R})\frac{m\theta\omega}{\hbar}-(\omega^{2}_{L}+\omega^{2}_{R})-\frac{m^2 \theta^2 \omega^{2}_{L}\omega^{2}_{R}}{\hbar^2}\right)\bar{z}(\omega)z(\omega) \right]
\label{15has}
\end{eqnarray}}}
{{where $P(\omega, \theta)=\frac{m\omega^2}{(1-\frac{m\omega\theta}{\hbar}+\frac{m^2 \theta^2 \omega^{2}_{R}}{\hbar^2})}$~. This motivates the choice of $K(\omega, \ell)$ in eq.(\ref{17}).}}

\noindent Substituting the form (\ref{11}) of $S_{I}$ in eq.(\ref{9}), we obtain the following flow equation for the coefficient $g(\omega, \ell)$:
{{
\begin{eqnarray}
\frac{\partial g(\omega, \ell)}{\partial\ell}=\left(-\frac{1}{\hbar\omega}+\frac{eB}{2\hbar m\omega^2}\right)g^{2}(\omega, \ell).
\label{19}
\end{eqnarray}}}
{{Note that we have ignored a vacuum term in the above equation which originates from the second term
on the right hand side of eq.(\ref{9}).}}

\noindent Integrating this equation subject to the initial condition (\ref{18}) leads to
\begin{eqnarray}
\label{ha7nn}
g(\omega, \ell)&=&\hbar\omega F(\omega, \ell)\\
F(\omega, \ell)&=&\frac{2m\omega(eB\omega +k)}{[(2m\omega-eB)(eB\omega +k)\ell-2\hbar m\omega^2]}
\label{ha7n}
\end{eqnarray}
{{which in turn implies (by dividing and multiplying the right hand side of eq.(\ref{ha7nn}) 
by $(1-\frac{m\omega\ell}{\hbar}+\frac{eB\ell}{2\hbar})$ and using eq.(\ref{17}))
\begin{eqnarray}
g(\omega, \ell)=\frac{\hbar\omega F(\omega, \ell)}{(1-\frac{m\omega\ell}{\hbar}+\frac{eB\ell}{2\hbar})}\left(1+\frac{eB\ell}{2\hbar}\right)
-K(\omega, \ell)F(\omega, \ell)\ell.
\label{ha7}
\end{eqnarray}}}
\noindent {{Using these results, rearranging the terms and rescaling the coordinates as 
$\tilde{z}(\omega)=\sqrt{1-F(\omega, \ell)\ell}~z(\omega)$, we obtain the effective action as
\begin{eqnarray}
S=\int d\omega\left[\bar{\tilde z}(\omega)K(\omega, \ell)\tilde{z}(\omega)
-\frac{(eB\omega+k)}{(1-\frac{m\omega\ell}{\hbar}+\frac{eB\ell}{2\hbar})[1+\frac{eB(eB\omega +k)\ell}{2\hbar m \omega^2}]}\left(1+\frac{eB\ell}{2\hbar}\right)
\bar{\tilde z}(\omega)\tilde z(\omega)\right]
+J_{\ell}[\tilde z, \bar{\tilde{z}}].
\label{ha8}
\end{eqnarray}
Making a further rescaling of coordinates as 
$\xi(\omega)=\frac{\tilde{z}(\omega)}{[1+\frac{eB(eB\omega +k)\ell}{2\hbar m \omega^2}]^{1/2}}$~,
the above form of the action can be recast as
\begin{eqnarray}
S=\int d\omega\left[\bar{\xi}(\omega)K(\omega, \ell)\xi(\omega)
-\frac{(eB\omega+k)}{(1-\frac{m\omega\ell}{\hbar}+\frac{eB\ell}{2\hbar})}
\bar{\xi}(\omega)\xi(\omega)\right]
+J_{\ell}[\xi, \bar{\xi}].
\label{ha8v}
\end{eqnarray}
The above action is found to be similar to the action of a particle in the presence of a (generalized) harmonic oscillator potential  moving in the noncommutative plane (\ref{zbas}). Indeed, considering the action (\ref{zbas}) rewritten in the Fourier space (\ref{15has}) (with $\theta$ being identified with the flow parameter $\ell$) and setting $\frac{m\theta\omega^{2}_{R}}{\hbar}=\frac{eB^{*}}{2m}$ and choosing $B^{*}=-\frac{2m^2 \omega_{L}^2 \theta}{e\hbar}$ leads to
\begin{eqnarray}
S&=&\int d\omega~\theta\left[\bar{z}(\omega)P(\omega, \theta)z(\omega)
-\frac{(eB^{*}\omega +k')}{(1-\frac{m\omega\theta}{\hbar}+\frac{eB^{*}\theta}{2\hbar})}\bar{z}(\omega)z(\omega) \right]
\label{15hasq}
\end{eqnarray}
where $P(\omega, \theta)=\frac{m\omega^2}{(1-\frac{m\omega\theta}{\hbar}+\frac{eB^{*}\theta}{2\hbar})}$ 
and $k'=m\left(\omega_{L}^2 +\omega_{R}^2 +\frac{eB^{*}\omega_{L}^2 \theta}{2\hbar}\right)$.
Comparing eq.(s)(\ref{ha8v}) and (\ref{15hasq}), clearly establishes the equivalence between the commutative and the noncommutative systems with the identification $B^{*}=B$ and $k'=k$.

\noindent The flows of the sources can also be obtained easily by substituting
eqs.(\ref{17}, \ref{ha7nn}, \ref{ha7n}) in eqs.(\ref{14a},\ref{14c}) which yields
\begin{eqnarray}
\label{ha11}
J_{1}(\ell)&=&J_{1}(0)\frac{2\hbar m\omega^2}{[(eB-2m\omega)(eB\omega+k)\ell +2\hbar m\omega^2]}\\
J_{0}(\ell)+J^{*}_{0}(\ell)&=&|J_{1}(0)|^{2}\frac{(eB-2m\omega)\ell}{[(eB-2m\omega)(eB\omega+k)\ell +2\hbar m\omega^2]}~.
\label{ha12uv}
\end{eqnarray}}}
The above equations together with eq.(\ref{ha8v}) constitutes a one parameter family of theories dual to the one of eq.(\ref{15ha}). 
The flow of the source terms establishes the precise dictionary between the correlation functions of the two systems.


\vskip 0.15cm
\noindent We now summarize our findings. In this paper, we have computed the action for the (generalized)
harmonic oscillator in a noncommutative plane employing the path integral approach developed in \cite{sgfgs}. 
The action for this system is derived in the coherent state basis with additional degrees of freedom.
From this the action in the coherent state basis without any additional degrees of freedom is then computed. This yields the
ground state spectrum of the system which agrees with that obtained using the canonical approach \cite{rohwer}. 
We also observe that the action looks similar to the action for a particle in a 
magnetic field in the presence of a harmonic oscillator potential moving in the noncommutative plane \cite{sgfgs1} 
but is not identical to it.
We then employ the exact renormalization group approach to show that an equivalence
can be constructed between this (noncommutative) system and a commutative system. This is done by flowing the commutative system
in a way which keeps the generating functional invariant. An important aspect of our method is the relaxation of the conditions imposed on the sources and the cutoff function. The duality gives a precise dictionary between the correlation functions of the noncommutative and commutative systems due to the flow of the source terms.

\vskip 0.07cm

{\bf{Acknowledgements}}

\noindent The author would like to thank the referees for very useful comments which helped in improving
the quality of the paper.


\end{document}